\documentclass[letter, amsfonts, amssymb, amsmath, reprint, showkeys, nofootinbib,superscriptaddress, twoside]{revtex4-1}
\usepackage[english]{babel}
\usepackage[utf8]{inputenc}
\usepackage{physics}
\usepackage{amsmath}
\usepackage{dsfont} 
\usepackage [english]{babel}
\usepackage [autostyle, english = american]{csquotes}
\MakeOuterQuote{"}
\DeclareMathOperator{\arccosh}{arccosh}

\usepackage[colorinlistoftodos, color=green!40, prependcaption]{todonotes}
\usepackage[pdftex, pdftitle={Article}, pdfauthor={Author}]{hyperref} 
\newcommand\ignore[1]{{}}
\begin{document}
\title{Boundary-induced topological transition in an open SSH model}


\author{Alexei Bissonnette}
\thanks{A. Bissonnette and N. Delnour contributed equally.}
\affiliation{Département de physique, Université de Montréal, Montréal, H3C 3J7, QC, Canada}
\author{Nicolas Delnour}
\email{Corresponding email: nicolas.delnour@mcgill.ca}
\affiliation{Department of Physics, McGill University, Montreal, H3A 2T8, QC, Canada }
\author{Andrew Mckenna}
\affiliation{Department of Physics, McGill University, Montreal, H3A 2T8, QC, Canada }
\author{Hichem Eleuch}
\affiliation{Department of Applied Physics and Astronomy, University of Sharjah, United Arab Emirates}
\author{Michael Hilke}
\affiliation{Department of Physics, McGill University, Montreal, H3A 2T8, QC, Canada }
\def\andname{} 
\author{Richard MacKenzie}
\affiliation{Département de physique, Université de Montréal, Montréal, H3C 3J7, QC, Canada}

\date{\today} 

\begin{abstract}
We consider a Su-Schrieffer-Heeger chain to which we attach a semi-infinite undimerized chain (lead) to both ends. We study the effect of the openness of the SSH model on its properties. A representation of the infinite system using an effective Hamiltonian allows us to examine its low-energy states in more detail. We show that, as one would expect, the topological edge states hybridize as the coupling between the systems is increased. As this coupling grows, these states are suppressed, while a new type of edge state emerges from the trivial topological phase. These new states, referred to as phase-inverted edge states, are localized low-energy modes very similar to the edge states of the topological phase. Interestingly, localization occurs on a new  shifted interface, moving from the first (and last) site to the second (and second to last) site. This suggests that the topology of the system is strongly affected by the leads, with three regimes of behavior. For very small coupling the system is in a well-defined topological phase; for very large coupling it is in the opposite phase; in the intermediate region, the system is in a transition regime.

\end{abstract}


\maketitle

\section{Introduction}
The Su-Schrieffer-Heeger (SSH) model is of great importance in physics due to the fact that it is one of the simplest models out of which emerge interesting topological properties such as edge states and solitons \cite{PhysRevResearch.4.013185,PhysRevB.96.205424,Zhang:21,Killian2022,Muoz2018TopologicalPO,ZAIMI2021127035}. The model is an excellent starting point to present important themes of condensed matter physics such as Bloch's theorem, chiral symmetries, adiabatic equivalents, topological invariants and the bulk-boundary correspondence \cite{2016}. It is also part of a group of systems that are of great interest: topological insulators \cite{article,unknownavik,articlebrandes,articlemcook}. These systems get their name from the insulating nature of their bulk, while also having conducting states on the boundaries stemming from a symmetry-protected topological order. One characteristic that makes these systems so unique is  the existence of low-energy states localized at the chain ends that are robust against disorder. As a consequence of its simplicity, many variations of the SSH model have been studied, including extensions to higher dimensions \cite{PhysRevResearch.4.023193,physics1010002,2019,PhysRevResearch.3.L042044} and generalizations including interactions, various forms of disorder, and driving \cite{Yahyavi_2018,PhysRevB.101.144204,PhysRevA.92.023624}.

In this paper, we study the properties of the SSH model in the case where the SSH chain is part of an open system. To do so, we attach the chain to a semi-infinite lead at each end; these leads reproduce the effect of an environment. Similar dissipative systems have been studied in the case where the effective potentials associated with the leads are taken to be constant (rather than being energy-dependent) \cite{OSTAHIE,2021,articleMarques,articleDangel}. This paper presents a generalization of these systems by including the energy dependence of the lead coupling on the SSH chain.

Our work brings several interesting phenomena to light. First, we illustrate how topological edge states can vanish as the coupling to the environment grows, while new low-energy states can emerge from the trivial phase. These new states bear a striking resemblance to edge states: their energies are near zero and they are localized on the edges of the SSH chain. Yet, there are significant differences; they appear when edge states do not exist in an isolated SSH chain. Furthermore, they are localized on the "wrong" sublattice (they present a strong localization on the second and second-to-last sites instead of the first and last sites). We refer to them as {\em phase-inverted edge states\/}, or {\em PIE states\/}. As we will see, they can be thought of as arising due to the decoupling of the first and last sites of the SSH chain as the coupling to the leads grows, leaving a chain with two fewer sites.

Second, we observe the appearance of states with energies beyond the SSH chain bands as the coupling to the environment is increased. These high-energy states, known as Tamm states \cite{tamm1932possible,tammpaper,PhysRevLett.101.113902}, are localized on the two pairs of sites that are linked by the lead coupling. We refer to these highly dimerized pairs of sites, composed of the first/last SSH site and the first site of the corresponding lead, as {\em islets\/}.

We then study the system through the lens of an effective description, incorporating the effect of the environment on the SSH chain in a modification of the SSH Hamiltonian. We argue that a topological phase transition occurs as the SSH-environment coupling becomes strong. This suggests a new way of modifying the topology of an SSH system without changing its internal structure: for a given initial topological phase, it is always possible to bring the system to the opposite phase. This concept can be extended to other one-dimensional topological insulators.  Finally, it is shown, as in the case of edge states, that PIE states are robust against certain types of disorder. We also show that they make significant and surprising contributions to the transport properties of the system, consistent with, and in support of, the other aspects of the paper.

The paper is organized as followed. In the next section, key elements which will be used throughout the article are introduced. A mathematical description of the tripartite system (SSH chain plus two leads) is given, along with its eigenvalues and eigenvectors. Special attention will be given to low-energy states and in particular the PIE states, after which Tamm states and the islets will be presented. In the third section, an effective description of the SSH subspace with the leads replaced by self-energies is analyzed. A comparison of this system in the limit of small and large coupling to the environment is presented. This will lead to the realization that a topological phase transition occurs between these two limits. In the fourth section, Green's functions are used to study the density of states, local density of states, the influence of disorder, and the transmission of the low-energy states when the coupling is neither very small nor very large. This will help us understand what happens between the two well-defined topological phases. In the last section,  we give a summary and comparison of the results obtained.

\section{The tripartite system}
\subsection{Description of the individual components}
\begin{figure}[h]
    \centering
\includegraphics[width=8.5cm]{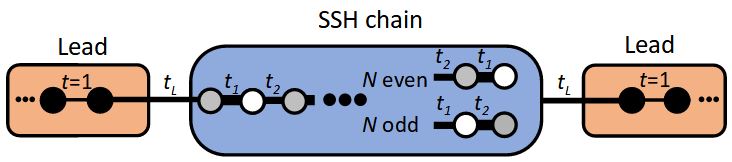}
    \caption{Visual representation of the tripartite system.}
    \label{f21}
\end{figure}
The tripartite system, which is represented in Figure \ref{f21}, has already been described in \cite{PhysRevA.95.062114, Nic1}. 
Its Hamiltonian takes the form \begin{equation}\label{eq1nic}
\begin{split}
H  &= \, H_{\text{SSH}}+H_{\text {L}}+H_{\text {R}} \\&
+t_L \big( c_1^\dagger l_1 + l_1^\dagger c_1 +c_N^\dagger r_1 + r_1^\dagger c_N\big),
\end{split}
\end{equation}
where \begin{align}\label{HSSH}
H_{\text{SSH}} = & \, \, \, \sum_{m=1}^{N-1} t_m \big( c^\dagger_m c_{m+1} + c_{m+1}^\dagger c_m\big),
\end{align}
\begin{equation}\label{Hsemiinf}
\begin{split}
H_{\text {L}} = & \sum_{m=1}^{\infty} \big( l_m^\dagger l_{m+1} +  l_{m+1}^\dagger l_{m} \big)\end{split},\end{equation}
and
\begin{equation}
\begin{split} H_{\text {R}} = & \sum_{m=1}^{\infty} \big( r_m^\dagger r_{m+1} +  r_{m+1}^\dagger r_{m} \big),
\end{split}
\end{equation}
where $m$ is the site index and $\{c_m^\dagger, c_m\}$, $\{l_m^\dagger,l_m\}$, $\{r_m^\dagger,r_m\}$ are the creation and annihilation operators for the $m^{\text{th}}$ site of the SSH chain, the left lead, and the right lead, respectively. 

The model is defined by the number of sites of the SSH chain $N$ and by the various hopping parameters, all assumed real and positive. We set the hopping parameter of the leads to unity (thus defining the energy scale). The strength of the coupling between the SSH chain and the leads is given by the hopping parameter $t_L$. The SSH chain hopping parameter $t_m$ (linking sites $m$ and $m+1$) alternates between two values, $t_1$ for $m$ odd and $t_2$ for $m$ even.

We remind the reader that the dispersion relation for the SSH chain is given by \cite{ZAIMI2021127035} \begin{equation}\label{dr1}
E^{2}=t_{1}^{2}+t_{2}^{2}+2 t_{1} t_{2} \cos 2 k ,
\end{equation}
while that for the leads is given by \begin{equation}\label{dr2}
E=2\cos(q).
\end{equation}
Here $k$ is the wave number in the SSH chain and $q$ is that in the leads.
We see from \eqref{dr1} that in the thermodynamic limit the SSH model has two symmetric bands bounded by $E=\pm (|t_1-t_2|,t_1+t_2)$; the gap disappears in the undimerized limit $t_1=t_2$. As we will see, for a finite SSH chain solutions exist within the chain for which the spatial behavior is oscillatory for energies within the bands and exponential outside the bands; these states are referred to as {\em bulk states\/} and {\em edge states\/}, respectively, reflecting their spatial profile (appreciable throughout the chain or localized at the ends). In both cases, whether such solutions exist for the tripartite system depends on the boundary conditions, here provided by the coupling to the leads.

The cases $N$ odd and even have some features in common as well as some differences. The most notable difference is already apparent for the well-studied case of an isolated SSH chain (our model in the limit  $t_L=0$). If $N$ is even the first and last coupling constants are $t_1$, whereas if $N$ is odd the last coupling constant is $t_2$ (see Fig.~\ref{f21}). This affects whether or not edge states exist, as can easily be surmised by considering the limiting case where one of the SSH hopping parameters is set to zero (see, for instance, \cite{2016}). In this limit, the chain breaks up into a number of uncoupled dimers internally linked by the nonzero hopping parameter and, possibly, one or two uncoupled sites at the ends of the chain. (Specifically, in the even case if $t_1=0$ there is an uncoupled site at each end whereas if $t_2=0$ there are none; in the odd case there is always one uncoupled site, on the left if $t_1=0$ and on the right if $t_2=0$.) In this limit the dimers have equal and opposite nonzero energies while the uncoupled sites have energy zero. As the coupling that had been set to zero is turned on, the former evolve to bulk states lying within the bands; the latter evolve to edge states. This suggests that for $N$ even, if $t_1<t_2$ there are two edge states of equal and opposite exponentially small energies whereas if $t_1>t_2$ no edge states exist. For $N$ odd, there will always be one zero-energy edge state, on the left or right for $t_1<t_2$ or $t_1>t_2$, respectively. These conclusions are valid, with one minor adjustment when $N$ is even: the transition from two edge states to none occurs not when $t_1=t_2$ but rather when the ratio of SSH hopping parameters $r\equiv t_1/t_2$ is at its critical value $r_\text{c}\equiv N/(N+2)$, which of course approaches unity as $N\to\infty$, as commonly seen in literature. Throughout the text, $r$ will be compared to unity for the sake of clarity, but $r_\text{c}$ is the true critical value and should  always be considered in the study of finite systems.

For the remainder of this article, we will take $N$ even, occasionally highlighting instances where the behavior for $N$ odd is significantly different.

\subsection{Solutions of the tripartite system}
Using Bloch's theorem \cite{Ashcroft76,dresselhaus2008applications} as an ansatz for the wave functions of the tripartite system, we obtain general expressions for the wave functions of each of its components: 

\begin{align}
\label{wfl1}
\ket{\psi_{\text{L}}}&=\sum_{n=1}^{\infty}{\big{(}e^{i(n-1)q}G_++e^{-i(n-1)q}G_-\big{)}\ket{n}}\nonumber\\
&\equiv\sum_{n=1}^{\infty}\psi_{\text{L},n}\ket{n},\\
\label{wfl2}
\ket{\psi_{\text{R}}}&=\sum_{n=1}^{\infty}{\big{(}e^{i(n-1)q}D_++e^{-i(n-1)q}D_-\big{)}\ket{n}}\nonumber\\
&\equiv\sum_{n=1}^{\infty}\psi_{\text{R},n}\ket{n},\\
\ket{\psi_{\mathrm{SSH}}}&=\sum_{m=0}^{N/2-1}\Big\{\big{(}C_+e^{-i\phi}e^{i2mk}+C_-e^{i\phi}e^{-i2mk}\big{)}\ket{2m+1}\nonumber\\&\pm \big{(}C_+e^{i\phi}e^{i2mk}+C_-e^{-i\phi}e^{-i2mk}\big{)}\ket{2m+2}\Big\}\nonumber\\
&\equiv\sum_{n=1}^{N}\psi_{\text{SSH},n}\ket{n}.
\label{eq-dog}\end{align}
Here, $\phi$ is a positive phase defined by $t_1+t_2e^{ 2ik}=|E|e^{2i\phi}$ for $E\neq0$. To avoid unnecessary clutter, we have labeled the site kets by their site number only and not by the component of the system (L, R or SSH). In \eqref{eq-dog}, the upper (lower) sign is for the positive (negative) energy solution.

The solutions (\ref{wfl1}-\ref{eq-dog}) must be combined such that the four boundary conditions (two at each interface) are satisfied. For $|E|<2$, all solutions in (\ref{wfl1},\ref{wfl2}) are valid, so there are six constants to be determined ($G_\pm,D_\pm,C_\pm$). With four boundary conditions and normalization, we see there are two independent solutions for any energy in this range. For $|E|>2$, from \eqref{dr2} we see that $q$ is imaginary and, choosing $\Im(q)>0$, the solutions $e^{-inq}$ in (\ref{wfl1},\ref{wfl2}) diverge; we must take $G_-=D_-=0$. The four boundary conditions and normalization overdetermine the four remaining coefficients, and only at certain discrete energies will solutions be found. 

In what follows, we will consider these energy domains separately. We begin, however, with a special case of the first case, namely $E=0$, which lends itself to a particularly simple analysis.

\subsubsection{Case 1: $E=0$}
As has just been argued, we expect two independent zero-energy eigenstates. Since for $E=0$ the Schrödinger equation links any site $n$ with sites $n\pm2$, the even and odd sublattices are decoupled from one another; thus the independent states can be taken to be nonzero on either the even or the odd sublattice. The general solution within the SSH chain, defined in terms of the hopping parameter ratio $r = t_1/t_2$, is given by
\begin{align}\label{fo0e}
\ket{\psi_{\text{SSH}}}=\alpha\begin{pmatrix}
    \begin{pmatrix}
    1\\
    0
    \end{pmatrix}\\
    \begin{pmatrix}
    1\\
    0
    \end{pmatrix}(-r)\\
    \vdots\\
    \begin{pmatrix}
    1\\
    0
    \end{pmatrix}(-r)^{\frac{N-4 }{2}}\\
    \begin{pmatrix}
    1\\
    0
    \end{pmatrix}(-r)^{\frac{N-2}{2}}\\
\end{pmatrix}+\beta\begin{pmatrix}
    \begin{pmatrix}
    0\\
    1
    \end{pmatrix}(-r)^{\frac{N-2}{2}}\\
    \begin{pmatrix}
    0\\
    1
    \end{pmatrix}(-r)^{\frac{N-4}{2}}\\
    \vdots\\
    \begin{pmatrix}
    0\\
    1
    \end{pmatrix}(-r)\\
    \begin{pmatrix}
    0\\
    1
    \end{pmatrix}
    \end{pmatrix}.
\end{align}
Note that the individual solutions, which we will refer to as the $\alpha$ and $\beta$ solutions, are parity inversions of one another, a consequence of the reflection invariance of the Hamiltonian \eqref{eq1nic}. Given this symmetry, in the rest of this section we will focus on the $\alpha$ solution; the same conclusions, inverted spatially, apply to the $\beta$ solution.

Let us take a more detailed look at the $\alpha$ solution, and specifically on its dependence on the hopping parameter ratio $r$. The factors $(-r)$ in \eqref{fo0e} give rise to an exponential attenuation (if $r<1$) or growth (if $r>1$) from left to right, with length scale $2|\log r|$. We see that if $r<1$ the solution is largest on the first site, very much like the edge states in an uncoupled SSH chain with $r<1$.

 If $r>1$ the largest amplitude is not on the {\em last\/} site but rather on the {\em second last\/} site. We thus observe that the solution behaves similarly to an edge state, with one important exception: it is strongly localized on the "wrong" sublattice compared to an edge state. For this reason, as mentioned above, we refer to this type of solution as a phase-inverted edge state, or PIE state.

Two points are worth noting for $r>1$. First, the behavior is completely unlike the uncoupled SSH chain, for which there are no states at all of energy near zero. Second, the PIE states bear a striking resemblance to the edge states of an isolated SSH chain of length $N-2$ beginning and ending with $t_2$, or in other words, of the SSH chain with its first and last sites removed.



How do these states extend beyond the SSH chain into the leads? The solutions in the leads are given by (\ref{wfl1},\ref{wfl2}) with $q=\pi/2$; these must be combined with \eqref{fo0e} such that the boundary conditions are satisfied. The eigenfunctions in the leads for the $\alpha$ solution turn out to be
\begin{equation}\label{xsolout}
\begin{aligned}
&\psi_{\text{L},n}=t_L\cos(\frac{n\pi}{2}),&\\
&\psi_{\text{R},n}=-\frac{t_1}{t_L}(-r)^{\frac{N-2}{2}}\sin(\frac{n\pi}{2}).
\end{aligned}
\end{equation}
Note that $\psi_{\text{L},n}$ is zero for $n$ odd while $\psi_{\text{R},n}$ is zero for $n$ even, maintaining the support on one sublattice throughout the system, as expected given the form of the Hamiltonian.

Eqs. \eqref{xsolout} provide a straightforward way to compare the amplitude of the wave functions within the SSH chain versus that in the leads for a given set of parameters $\{t_1, t_2, t_L\}$. Since the solutions are exponential in the SSH chain and oscillatory in the leads, it's simply a matter of comparing the first (last) nonzero amplitude of the SSH chain given in \eqref{fo0e} with the amplitudes given in \eqref{xsolout}. In particular, we find
\begin{equation}\label{xsolout2}
\begin{aligned}
&|\psi_{\text{L},2}|=t_L |\psi_{\text{SSH},1}|,&\\
&|\psi_{\text{R},1}|=\frac{t_1}{t_L}|\psi_{\text{SSH},N-1}|.
\end{aligned}
\end{equation}
On the left boundary, if $t_L>1$, the amplitude of the first lead site with nonzero amplitude is larger than the first site of the SSH chain, and if $t_L<1$, the opposite is true. On the right boundary, if $t_1>t_L$, the amplitude of the first site of the lead with nonzero amplitude (the second site) is larger than the first site of the SSH chain, and if $t_1<t_L$, the opposite is true. Interestingly, this conclusion doesn't depend on whether the zero-energy state is an edge state or a PIE state; that is to say, whether $r>1$ or $r<1$, the zero-energy $\alpha$-state amplitude going from the left edge of the SSH chain to the lead is multiplied by $t_L$ whereas on the right side it is multiplied by $t_1/t_L$. These multiplicative factors will reappear in Section \ref{sec-dog} where an effective description of the system is discussed.

\begin{figure}
    \centering
    \includegraphics[width=7.5cm]{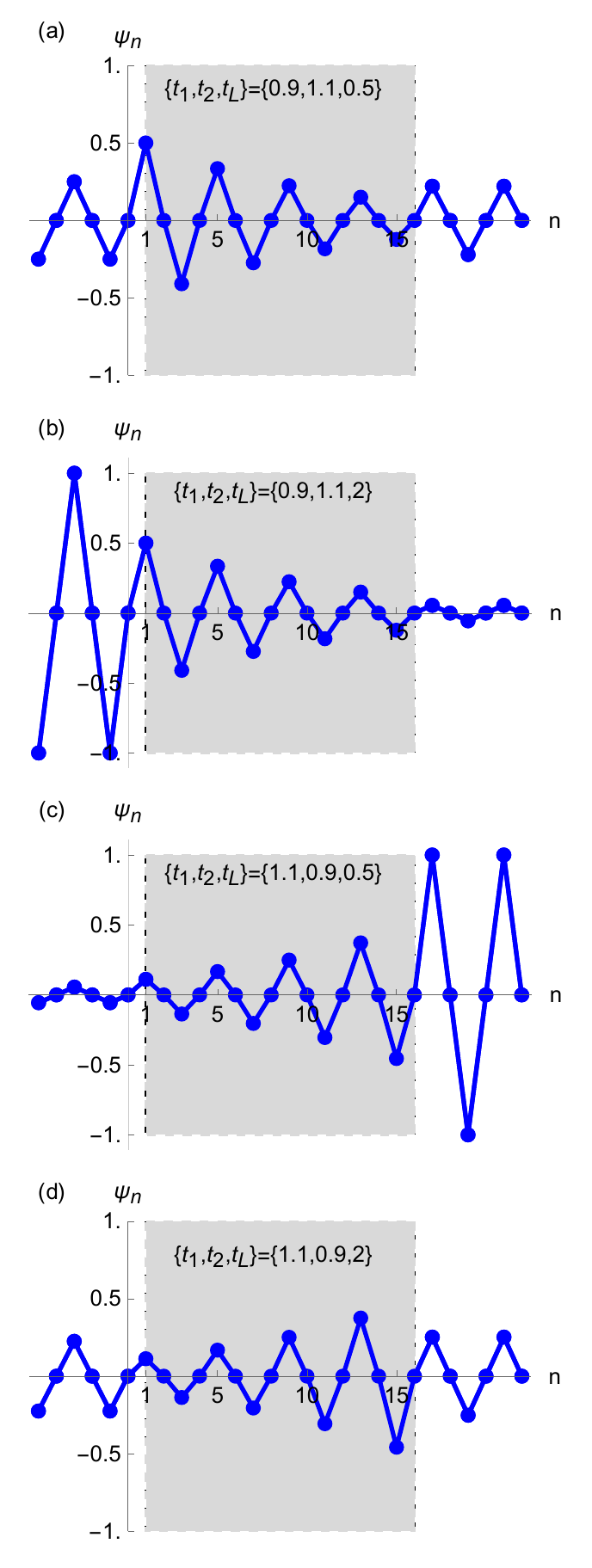}
    \caption{Wave functions of the zero-energy $\alpha$ state of an SSH chain of length $N=16$. (The $\beta$ state is identical up to spatial reflection.) The SSH chain (shaded area) and the first 6 sites of both leads (outside the shaded area) are displayed for four sets of parameters. In a) and b) $t_1<t_2$ so the state is a left edge state; in c) and d) $t_1>t_2$ so the state is a right PIE state. In all cases, he amplitude of the oscillating wave function in the leads compared with that in the SSH chain behaves in accordance with \eqref{xsolout2}.
    }
    \label{figLESSS}
\end{figure}

We illustrate these zero-energy $\alpha$ states, and in particular the amplitude factors \eqref{xsolout2}, for different values of $t_1, t_2, t_L$ in Figure \ref{figLESSS}.

We conclude this section with a brief summary of the properties of zero-energy states if $N$ is odd. Since the first and last coupling constants are $t_1$ and $t_2$, respectively, spatial inversion is still a symmetry when combined with the change $t_1\leftrightarrow t_2$. As mentioned earlier, an isolated odd-$N$ SSH chain has one zero-energy edge state, on the left if $t_1<t_2$ and on the right if $t_1>t_2$. When the leads are attached the edge state remains an edge state; in addition, there is a second solution which is a PIE state localized on the other end of the chain. The solution within the SSH chain is given by \eqref{fo0e} with $N\to N-1$ and an added component at the end. (Thus, the final component of the $\alpha$ solution is $(-r)^{(N-1)/2}$ while that of the $\beta$ solution vanishes.) Whereas for $N$ even spatial inversion exchanges the $\alpha$ and $\beta$ solutions, for $N$ odd spatial inversion combined with $t_1\leftrightarrow t_2$ maps the $\alpha$ solution with $r>1$ to that with $r<1$ and vice-versa, and similarly for the $\beta$ solution. For the $N$-odd $\alpha$ solution, \eqref{xsolout} becomes
\begin{equation}\label{xsolout1NoddAlpha}
\begin{aligned}
&\psi_{\text{L},n}=t_L\cos(\frac{n\pi}{2}),&\\
&\psi_{\text{R},n}=t_L(-r)^{\frac{N-1}{2}}\cos(\frac{n\pi}{2}).
\end{aligned}
\end{equation}
and \eqref{xsolout2} becomes
\begin{equation}\label{xsolout2NoddAlpha}
\begin{aligned}
&|\psi_{\text{L},2}|=t_L |\psi_{\text{SSH},1}|,&\\
&|\psi_{\text{R},2}|=t_L|\psi_{\text{SSH},N}|.
\end{aligned}
\end{equation}
The equivalent equations for the $N$-odd $\beta$ solution are
\begin{equation}\label{xsolout1NoddBeta}
\begin{aligned}
&\psi_{\text{L},n}=-\frac{t_1}{t_L}(-r)^{\frac{N-3}{2}}\sin(\frac{n\pi}{2}),&\\
&\psi_{\text{R},n}=-\frac{t_2}{t_L}\sin(\frac{n\pi}{2})
\end{aligned}
\end{equation}
and
\begin{equation}\label{xsolout2NoddBeta}
\begin{aligned}
&|\psi_{\text{L},1}|=\frac{t_1}{t_L} |\psi_{\text{SSH},2}|,&\\
&|\psi_{\text{R},1}|=\frac{t_2}{t_L}|\psi_{\text{SSH},N-1}|.
\end{aligned}
\end{equation}
Notice that the multiplicative factors in \eqref{xsolout2NoddAlpha} and \eqref{xsolout2NoddBeta} are the expected generalization of those in \eqref{xsolout2}.

\subsubsection{Case 2: $|E|<2\neq 0$}
As argued earlier, when $|E|<2$ the wave number $q$ is real and there are two solutions for any energy. The solutions within the SSH chain are given, up to normalization, by 
\ignore{
\begin{equation}\label{jj2}
\begin{split}
&\Psi_{\text{SSH},n}=\\&
\left\{
    \begin{array}{llll}
    e^{iq}\big{(}(Et_L^2e^{-iq}-t_2^2)s_{n-1}-t_1t_2s_{n+1}\big{)}\\
~~+\alpha\,e^{-iq}\big{(}
    t_1t_L^2e^{iq}s_{N-n-1}-t_2(E-t_L^2e^{2iq})s_{N-n+1}\big{)}\\
\qquad\qquad(n\text{ odd}),
\\
\\e^{iq}\big{(}t_1t_L^2e^{-iq}s_{n-2}-t_2(E-t_L^2e^{-iq})s_n\big{)}\\
    ~~+\alpha\,e^{-iq}\big{(}
    (Et_L^2e^{iq}-t_2^2)s_{N-n}-t_1t_2s_{N-n+2}\big{)}\\
\qquad\qquad(n\text{ even}), 
    \end{array}
\right.
\end{split}
\end{equation}
}
\begin{equation}\label{jj2}
\begin{split}
&\Psi_{\text{SSH},n}=\\&
\left\{
    \begin{split}
    &\alpha \left([Et_L^2e^{-iq}-t_2^2]s_{n-1}-t_1t_2s_{n+1}\right)\\
&~~+\beta\,\left(
    t_1t_L^2e^{iq}s_{N-n-1}-t_2[E-t_L^2e^{iq}]s_{N-n+1}\right)\\
&\qquad\qquad(n\text{ odd}),
\\
&\alpha \left(t_1t_L^2e^{-iq}s_{n-2}-t_2[E-t_L^2e^{-iq}]s_n\right)
\\
    &~~+\beta\,\left(
    [Et_L^2e^{iq}-t_2^2]s_{N-n}-t_1t_2s_{N-n+2}\right)\\
&\qquad\qquad(n\text{ even}), 
    \end{split}
\right.
\end{split}
\end{equation}
where $s_n=\sin(nk)$ and
$\alpha$ and $\beta$ are arbitrary parameters. Note that, as with \eqref{fo0e}, the solutions are parity inversions of one another, under the combined operation $n\to N-n+1$, $q\to-q$. For energies in the range $|t_1-t_2|\leq |E| \leq |t_1+t_2|$ the wave vector $k$ is real and the wave functions in the SSH chain are oscillatory, whereas for energies outside these bands $k$ is complex and the wave functions are exponential. 

\subsubsection{Case 3: $|E|>2$}
\begin{figure}
    \centering
\includegraphics[width=8cm]{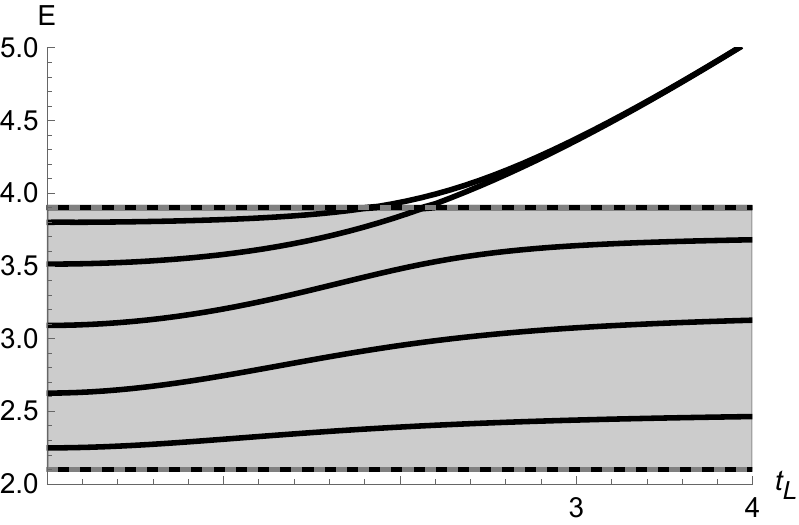}
    \caption{Graph representing the energy solutions of (\ref{eqt1}) as a function of $t_L$ with $\{t_1, t_2, N\}=\{3, 0.9, 10\}$ for positive energy values outside of the lead band. The shaded area is the positive-energy band of the SSH chain. We observe, in particular, that two states leave this band: these are known as Tamm states. $t_1$ and $t_2$ were chosen so that the SSH bands do not overlap the lead band. Thus, there is a maximal number of states with energy $|E|>2$. Note also that the behavior for $r<1$ is qualitatively similar to the displayed graph, for which $r>1$.
    \label{spec_1}
    }
\end{figure}

From (\ref{dr2}), we note that any state with energy $|E|>2$ has a complex wave number $q$ in the leads, which from (\ref{wfl1}) and (\ref{wfl2}) implies that the wave functions in the leads are given by a linear combination of states whose amplitudes increase and decrease exponentially with distance from the SSH chain. Since the leads are infinite, the exponentially increasing states are unphysical, so $G_-$ and $D_-$ must be set to zero (with the choice $\Im(q)>0$). Having four boundary conditions (two per boundary), one normalization condition, and five unknowns ($C_\pm$, $G_+$, $D_+$ and $E$), making use of the dispersion relations (\ref{dr1},\ref{dr2}) for $q$ and $k$, there are a discrete set of energies, given by the solutions of the following transcendental  equation:
\begin{multline}
t_1\left(t_2^2e^{-iq}s_{N+2}+t_L^4e^{iq}s_{N-2}\right)\\
+t_2\left(t_2^2e^{-iq}-2Et_L^2+t_L^4e^{iq}\right)s_N=0.
\label{eqt1}
\end{multline}
This equation can be solved numerically for a given set of parameters $\{t_1, t_2, t_L, N\}$. For fixed $t_1$, $t_2$ and $N$, the energy spectrum of the tripartite system is a function of $t_L$. In particular, if $|t_1-t_2|>2$, all bulk states of the SSH chain have $|E|>2$, as shown in Figure \ref{spec_1}. The corresponding wave functions are given, up to normalization, by
\begin{equation}\label{8b}\begin{aligned}
&\psi_{\text{SSH},n}=
\left\{
    \begin{array}{ll}
        t_1t_2s_{n+1}+(t_2^2-Et_L^2e^{iq})s_{n-1} & (n\text{ odd}),   \\
        t_2(E-t_L^2e^{iq})s_n-t_1t_L^2e^{iq}s_{n-2} & (n\text{ even}).
    \end{array}
\right.
\end{aligned}
\end{equation}
Note that $q$, given by \eqref{dr2}, is complex here. More precisely, $q=i\arccosh(E/2)$ for $E>2$ and $q=\pi+i\arccosh(|E|/2)$ for $E<-2$.

From Figure \ref{spec_1}, we observe that as $t_L$ is increased, two states leave the upper energy band of the SSH chain. These surface states have energies outside the SSH bands, $|E|>t_1+t_2$, and are known as Tamm states in the tight-binding formalism \cite{tamm1932possible}. Note that there are no equivalent states for the closed SSH system: these states are a result of the coupling between the SSH chain and the environment. For such states, we see from (\ref{dr1}) that the wave number $k$ in the SSH chain is complex, meaning that the corresponding wave functions have exponential-like behaviors. While these states may appear similar to edge states due to their localization at the boundaries, they are high-energy excitations which show no sublattice confinement.
\begin{figure}
    \centering
\includegraphics[width=8.5cm]{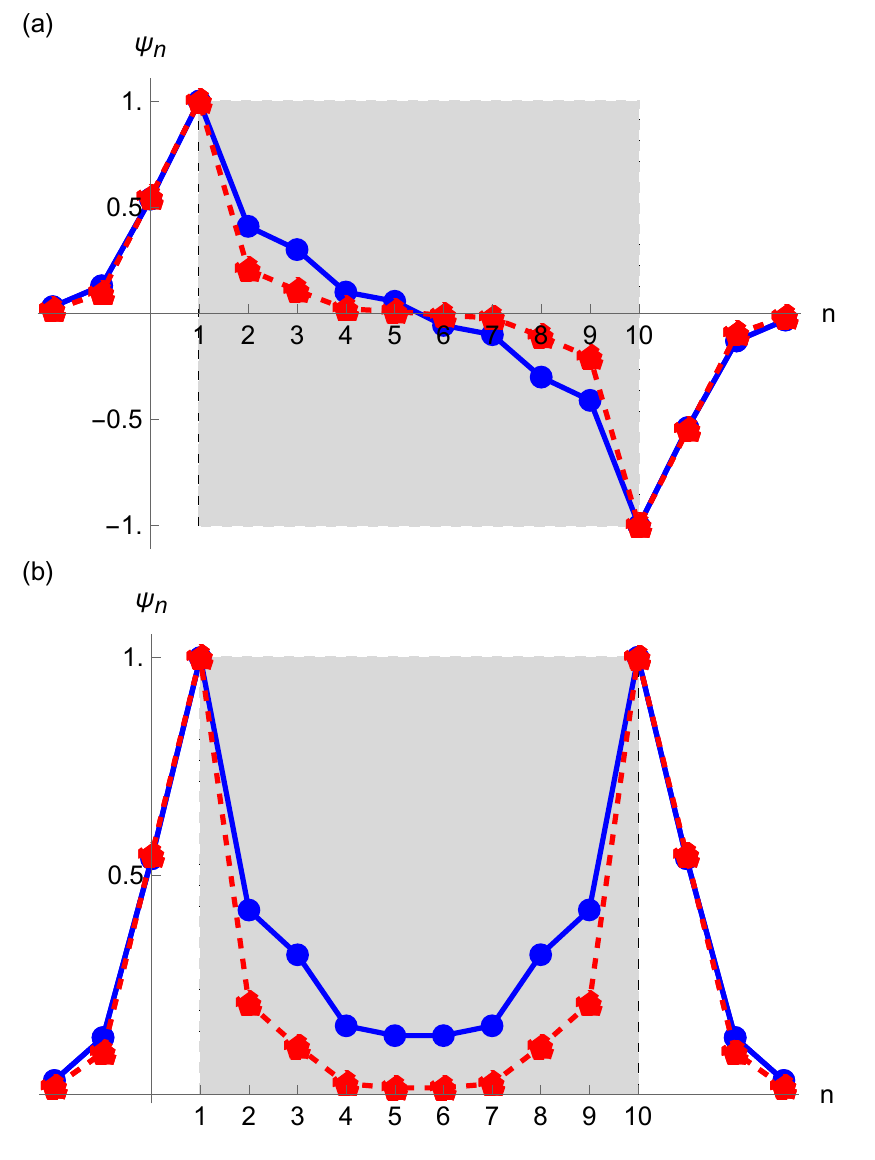}   
    \caption{High-energy Tamm states beyond the SSH band for $\{t_1, t_2, N\}=\{3, 0.9, 10\}$. We present the wave functions associated with the SSH chain (in the shaded area) and the wave functions associated with the first three sites of both leads (outside the shaded area). These states are strongly localized on the boundaries and have exponentially decreasing profiles as they enter the bulk of both the SSH chain and the leads. Note that this behavior is accentuated as $t_L$ is increased. (a) Antisymmetric Tamm state, with $t_L = 3$ ($E= 4.39$) for the blue (solid) curve and $t_L=5$ ($E=5.92$) for the red (dashed) curve. (b) Symmetric Tamm state, with $t_L=3$ ($E=4.40$) for the blue (solid) curve and $t_L=5$ ($E=5.92$) for the red (dashed) curve. In both cases, we clearly see that the Tamm states associated with the larger coupling constant $t_L$ is more strongly located on the boundaries of the SSH chain. In the limit $t_L\rightarrow \infty$, only sites 0, 1, 10 and 11 will have nonzero amplitudes.
}
    \label{figHES}
\end{figure}
In Figure \ref{figHES}, we display these states for a given set of $\{t_1, t_2, N\}$, and for two different values of $t_L$. This allows us to see how the Tamm states are affected by the coupling parameter $t_L$: the larger this parameter is, the more the states are confined to the boundaries. In the limit of $t_L\rightarrow \infty$, the wave function of the tripartite system will have a nonzero amplitude only on the two pairs of sites associated with the two islets that will be introduced shortly. In parallel, going back to Figure \ref{spec_1}, we observe that the remaining states asymptotically approach fixed energies within the SSH bands. Thus, depending on $t_L$, we observe either zero, two, or four Tamm states.

From (\ref{eq1nic}), we can see how these Tamm states emerge. When $t_L$ is much larger than the other couplings, the two states within an islet couple to each other much more strongly than to the adjacent sites, effectively decoupling them from the rest of the system; they are then described by the Hamiltonian \begin{equation}\label{hamisland}
H_\text{islet}=t_L\sigma_1 = \left(\begin{array}{cc}
0&t_L\\
t_L&0
\end{array}\right),
\end{equation}
with eigenvalues $E=\pm t_L$, which give the energies of the Tamm states in the limit of strong coupling $t_L \gg t_1, t_2, 1$. See the Appendix
for a more complete discussion of the strong coupling case.

\begin{figure}
    \centering
    \includegraphics[width=8.5cm]{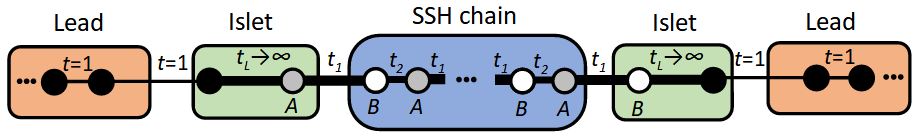}
    \caption{Tripartite system in the limit of strong coupling. The SSH chain loses its first and last site to the islets. Thus, its first and last hopping parameters are both $t_2$ in this limit.
    }
    \label{islets}
\end{figure}

This leads us to a different perspective of the tripartite system when $t_L$ is large. In this limit, two islets, described by the Hamiltonian (\ref{hamisland}), emerge between the three components of the system. Figure \ref{islets} illustrates the system in this limit. In the limit $t_L\rightarrow \infty$, the five components (two  leads, two islets, SSH chain), are disconnected from one another. In particular, we observe that the "final" uncoupled SSH chain is of length $N-2$: it has lost the two sites that were connected to the leads, giving rise to a topological phase transition that will be discussed in Section 3.

Studying the tripartite system leads to many insights, yet it leaves several questions unanswered. In particular, it doesn't explain how the low-energy states are affected by the environment. In the next section, this will be clarified using effective potentials, a formalism that incorporates the effect of the leads in a modification of the SSH Hamiltonian.

\section{Effective representation of the tripartite system}
\begin{figure}[h]
    \centering
    \includegraphics[width=8cm]{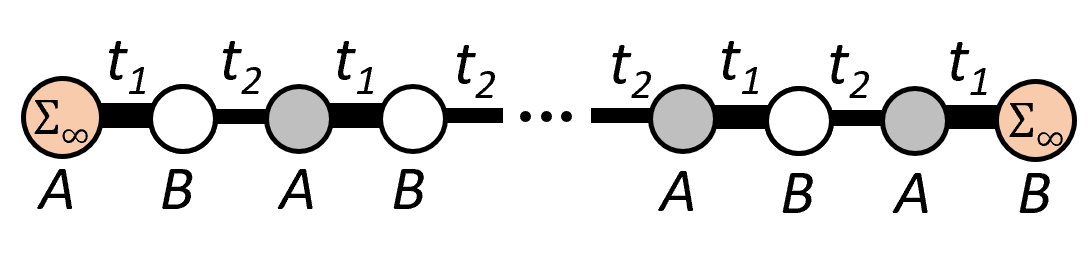}
    \caption{Graphic representation of the effective system. The effective potentials $\Sigma_\infty$ on the first and last sites represent the effect of the leads on the SSH chain.}
    \label{fig:my_label}
\end{figure}
The tripartite system can be written in the form of an $N \times N$ effective Hamiltonian for the subspace of the SSH chain, where leads have been replaced by self-energies. It has been shown in \cite{PhysRevA.95.062114, ZAIMI2021127035, Nic1} that the  effective representation of this system is given by \begin{equation}\label{Heff}
H_{\text{SSH}}^{\text{eff}}=H_{\text{SSH}}+\Sigma_\infty(E) (c_1^{\dagger}c_1+c_N^{\dagger}c_N),
\end{equation}
where 
\begin{equation}\label{exptl}
\Sigma_\infty(E) = t_L^2 e^{-iq}
\end{equation}
is the self-energy of a lead, taking the form a complex energy-dependent potential. To avoid phase discontinuity issues in the self-energy, a dissipative causal condition,  $\Im{\Sigma_\infty}\leq0$, is imposed on the self-energy by way of a branch cut orientation change. The conventional branch cuts of $\Sigma_\infty$ along the real-$E$ axis from $[-2,2]$ are  rotated about $\Re{E}=\pm2$ to extend towards $\Im{E}<0$, which reduces (\ref{exptl}) to
\begin{equation}
\label{expt2}
\begin{aligned}
\Sigma_\infty = \frac{t_L^2}{2} \left\{
    \begin{array}{ll}
       E-i\sqrt{4-E^2} & \mbox{for $|\Re{E}|\leq 2$,}\vspace{0.1cm}   \\ 
       E-\text{sgn}(\Re{E})\sqrt{E^2-4} & \mbox{for $|\Re{E}|> 2$,}
    \end{array}
\right.
\end{aligned}
\end{equation}
where for $E\notin \mathbb{R}$ the square root is taken to have a positive real part. This form for $\Sigma_\infty$ is equivalent to the conventional self-energy of semi-infinite leads \cite{datta_2005} along the real-$E$ axis, and thus yields identical physical quantities. Rotating the branch cuts this way is useful, as it reveals the dissipative poles associated with the complex solutions of (\ref{Heff}).

In contrast to the full Hamiltonian \eqref{eq1nic}, (\ref{Heff}) is non-Hermitian. Because of this, the energies of the effective system will in general be complex, leading to non-conservation of probability. Although strange at first sight, this simply reflects the fact that the effective representation includes only a finite spatial segment of the total tripartite system. Real parts of the energy, $\Re{E}$, then correspond to the \textit{physical} energies of states supported within the SSH chain subspace, while the imaginary parts, $\Im{E}$, are related to the decay of states out of that subspace. Equivalently, the imaginary components describe hybridization and the coupling of the subspace to its environment, leading to a broadening of the energy spectrum.

\subsection{$N$-site effective system}
Solving the Schrödinger equation for the Hamiltonian given in (\ref{Heff}), one can use either the right or left boundary conditions of the system to find an expression for the wave functions. Using the left boundary condition, the unnormalized wave functions are found to be \begin{equation}\label{x9}
\begin{aligned}
&\ket{\psi_{n}}=\mp\frac{ 2ie^{i\phi}C_+}{t_2Ee^{2ik}-\Sigma_\infty(t_1+t_2e^{2ik})}&\\&\times
\left\{
    \begin{array}{ll}
       (E\Sigma_\infty-t_2^2)\text{s}_{n-1}-t_1t_2\text{s}_{n+1} & \mbox{if $n$ is odd,}   \\
        t_1\Sigma_\infty \text{s}_{n-2}-t_2(E-\Sigma_\infty)\text{s}_n & \mbox{if $n$ is even.}
    \end{array}
\right.
\end{aligned}
\end{equation}
where as mentioned earlier $\phi$ is defined by $t_1+t_2e^{2ik}=|E|e^{2i\phi}$, and where we keep the multiplicative factor for a later calculation.
Using the right boundary condition, the unnormalized wave functions are given by
\begin{equation}\label{x10}
\begin{aligned}
&\ket{\psi_{n}}=\frac{ 2ie^{-i\phi}C_+e^{i(N-2)k}}{t_2Ee^{-2ik}-\Sigma_\infty(t_1+t_2e^{-i2k})}&\\&\times
\left\{
    \begin{array}{ll}
       t_1\Sigma_\infty \text{s}_{N-n-1}-t_2(E-\Sigma_\infty )\text{s}_{N-n+1} & \mbox{if $n$ is odd,}   \\
        (E\Sigma_\infty -t_2^2)\text{s}_{N-n}-t_1t_2\text{s}_{N-n+2} & \mbox{if $n$ is even.}
    \end{array}
\right.
\end{aligned}
\end{equation}
Note that (\ref{x9}) and (\ref{x10}) are proportional to the $\alpha$ and $\beta$ terms of (\ref{jj2}), respectively. However, the effective system is more constrained than the tripartite system, and rather than two independent solutions with a continuum of energies for $|E|<2$, we expect a single solution forming a discrete set of energies. Thus, (\ref{x9}) and (\ref{x10}) must in fact agree, leading to the following transcendental equation for the energy: \begin{align}\label{x12}
\begin{split}
&t_1(t_2^2\text{s}_{N+2}+\Sigma_{\infty}^2\text{s}_{N-2})\\
&+t_2(t_2^2-2E\Sigma_{\infty}+\Sigma_{\infty}^2)\text{s}_N=0.
\end{split}
\end{align}
For $|\Re{E}|>2$, \eqref{eqt1} and (\ref{x12}) agree, as they must.

The domain of real values of $E$ determines what type of solutions arise in the effective system. Solutions with $|\Re{E}|>2$ existing outside of the lead bands are immune to hybridization and must be real-energy solutions, as confirmed by $\Im{\Sigma_\infty}=0$ along the real-axis. These states therefore exhibit no broadening. On the other hand, if $|\Re{E}|<2$ then $\Sigma_\infty$ is complex, resulting in complex-energy  solutions to (\ref{Heff}) and (\ref{x12}). This is to be expected, since states in the SSH subspace with energies $|\Re{E}|<2$ are coupled to the lead bands; they hybridize and feature a broadened spectrum within this subspace. 
\begin{figure}[h]
    \centering
    \includegraphics[width=1\linewidth]{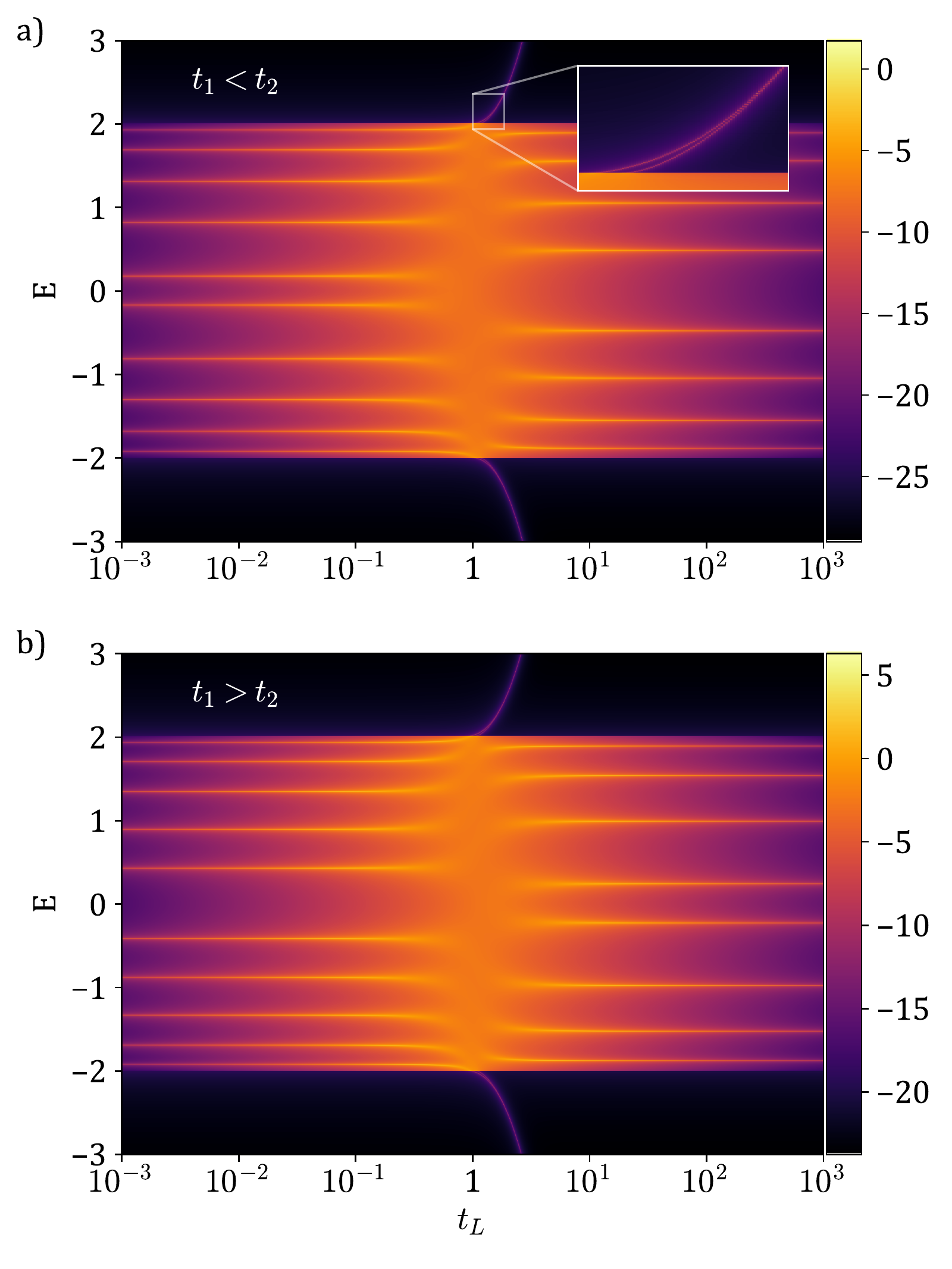}
    \caption{
    Graph of the log of the density of states (see section IV) for real values of $E$ in the lead band, as a function of $t_L$, with a) $\{t_1, t_2, N\}=\{0.9, 1.1, 10\}$, b) $\{t_1, t_2, N\}=\{1.1, 0.9, 10\}$. The points $\{E, t_L\}$ associated with large numerical values (see the color code) represent states for which the imaginary component of the energy is small (the projection of these states on the real axis is closer to being a singularity of \eqref{x12}). We also observe that the energy spectrum is broadened: this testifies to the fact that the energies are complex in this region ($|E|<2$). The positions of the maxima correspond to the real part of the solution in \eqref{x12}, while the imaginary part of the solution is inversely proportional to the density of states along the maxima (bright colors).}
    \label{figspp}
\end{figure}

Broadening is observed in Fig.~\ref{figspp}a) (conventional edge states) and Fig.~\ref{figspp}b) (PIE states), which plots the density of states (see section IV), whose maxima are inversely proportional to the imaginary part of the solutions of (\ref{x12}) within the lead bands.
For conventional edge states at small $t_L$, the spectrum shows minimal broadening: one clearly observes ten solutions corresponding to the eigenstates of an isolated SSH chain of length $N=10$. As $t_L$ approaches unity, the broadening increases, and the spectrum reaches a point where it is maximally broadened and continuous. Beyond $t_L\approx1$, two pairs of Tamm states leave the energy band of the leads, consistent with the vanishing edge states and the SSH chain losing two sites to the islets for increasing coupling strength $t_L$. In the limit $t_L\rightarrow\infty$, the spectrum sharpens and  becomes discretized, corresponding to the spectrum of an isolated SSH chain of length $N-2$ beginning and ending with $t_2$. In this limit, the topological edge states present for small $t_L$ have disappeared. Note that for $t_L=1$, the coupling constant is equal to the hopping parameter in the leads, meaning that the leads gain two sites and the SSH chain has as first hopping parameter $t_2$ instead of $t_1$, and a lead coupling $t_1$ instead of $t_L$. This maximizes hybridization between the SSH states and the leads' states, which favors the  decay of the SSH states into the leads and explains maximal broadening is observed close to $t_L=1$. 

The case of PIE states, shown in Fig.~\ref{figspp}b, shares many features with the  conventional SSH chain system. Namely, the influence of broadening on the DOS is exerted identically, and Tamm states also arise for $t_L>1$. An important difference is the low-energy mode behavior. For small $t_L$, the SSH chain has $r>1$ and therefore hosts no edge states. For $t_L$ values larger than 1, the SSH chain can now be understood of having an effective length $N-2$, and hence finds itself closer to a nontrivial topological configuration. Low-energy modes, the so-called PIE states, are seen to emerge, becoming discretized as $t_L\rightarrow \infty$.

\subsection{$(N-2)$-site effective system}\label{sec-dog}
To understand the emergence of PIE states, consider an $(N-2)$-site effective system, given by \begin{equation}\label{y4}
H_{\text{SSH}'}^{\text{eff}}=H_{\text{SSH}'}+\sigma_\infty(E) (c_2^{\dagger}c_2+c_{N-1}^{\dagger}c_{N-1}),
\end{equation}
where $H_{\text{SSH}'}$ is the Hamiltonian of an SSH chain of length $N-2$ whose first and last hopping parameters are $t_2$; furthermore, $\sigma_\infty(E)=t_1^2/(E-\Sigma_\infty(E))$. In this case, the effective potential $\sigma_{\infty}$ describes the effect of the leads and the first and last sites of the SSH chain on the remainder of the SSH chain. Note that as $t_L$ goes to infinity, $\sigma_\infty$ goes to zero and $H_{\text{SSH}'}^{\text{eff}}$ reduces to $H_{\text{SSH}'}$. We see explicitly that the islets and leads decouple from the reduced SSH chain, as mentioned earlier.

Thus, for large coupling, a \textit{new} effective SSH system emerges, with first and last coupling parameters $t_2$ rather than $t_1$, which is significant. If initially $r<1$, then the new SSH system has $r>1$, and vice-versa. In other words, a system that initially had two edge states will have no edge states for large coupling, and a system that initially had no edge states will have two PIE states for large coupling. This is a clear sign that a topological phase transition occurs between these two limits: for all values of $t_1$ and $t_2$, a state with a given topological phase at $t_L=0$ will be found in the opposite phase for $t_L\rightarrow \infty$.

If $N$ is odd, the same argument cannot be made. In this case, the system  hosts both edge and PIE states for any $r\neq 1$ and finite non-zero $t_L$ and therefore always shows low-energy topological signatures. Only in the limit $t_L\rightarrow0$ ($t_L\rightarrow\infty$) can edge states (PIE states) become the unique low-energy solutions within the effective system. The crossover in the vicinity of $t_L=1$ is accompanied by a delocalization-localization transition of low-energy modes; states initially  on the left (right) SSH chain edge are displaced to the right (left) edge as $t_L$ transitions through $1$. Unlike the isolated odd-$N$ SSH chain, the odd-length chain here hosts low-energy topological modes with support near both boundaries.

We conclude this section with a brief remark making a connection between the multiplicative factors appearing in (\ref{xsolout2},\ref{xsolout2NoddAlpha},\ref{xsolout2NoddBeta}) and the self-energies appearing in the effective Hamiltonians \eqref{Heff} and \eqref{y4}. The factors in (\ref{xsolout2},\ref{xsolout2NoddAlpha},\ref{xsolout2NoddBeta}) multiply the last nonzero amplitude in the SSH chain to give the amplitude of the oscillating wave function in the lead (for the specific case of zero-energy solutions). It was seen that in the case of an edge state, for which the last nonzero amplitude in the SSH chain is on the first or last site, the multiplication factor is $t_L$ (first equation in \eqref{xsolout2}, both equations in \eqref{xsolout2NoddAlpha}), whereas in the case of a PIE state, for which the last nonzero amplitude is on the second or second last site, the multiplication factor is $t_1/t_L$ (second equation in \eqref{xsolout2}, both equations in \eqref{xsolout2NoddBeta}). These factors, when squared, correspond exactly to the self-energies (evaluated at $E=0$) of the effective Hamiltonian if the leads, or the leads plus the first and last SSH sites, are integrated out; that is, they correspond to $|\Sigma_\infty(E=0)|=t_L^2$ and $|\sigma_\infty(E=0)|=(t_1/t_L)^2$, respectively. Thus, the self-energies appearing in the effective Hamiltonians not only give the effect of the part of the system that is integrated out on the system that remains; they also give the amplitude of the wave function in the part of the system that is integrated out in terms of the wave function in the system that remains.

\section{Green's function of the effective system}
In this section, the aim is to further understand the behavior of low-energy states between the weak- and strong-coupling limits, $t_L\to0$ and $t_L\rightarrow\infty$, respectively. As we have already seen, for $r<1$ the low-energy modes are edge states whereas for $r>1$ the low-energy modes are PIE states. While the effective system admits a continuum of energies due to hybridization with the leads, the edge states and PIE states have nonzero energies in finite systems, and are referred to as low-energy modes. 

\subsection{Density of States}
The DOS and the local density of states (LDOS) will prove to be key quantities in understanding the $N$-site effective system's response to changes of $t_L$. The DOS for an energy $E$ can be readily found by taking the trace of the imaginary part of the effective system's Green's function:
\begin{equation}\label{dosgreens}
    D(E)=-\frac{1}{\pi}\Im \{\mathrm{Tr}\,G^{\,\mathrm{eff}}_{\text{SSH}}(E)\}.
\end{equation}
It is a straightforward calculation to obtain such a Green's function, $G^{\,\mathrm{eff}}_{\mathrm{SSH}}=\left( E - H_{\mathrm{SSH}}^{\mathrm{eff}} \right)^{-1}$, which depends on the parity of $N$ and $n$.
For effective systems where $N$ is even, the diagonal elements of the Green's function $\left(G_{\mathrm{SSH}}^{\,\mathrm{eff}}\right)_{nn}$ were previously derived and documented in \cite{Nic1}. The DOS obeys a sum rule stating that an integrated peak profile for a single state must sum to 1 \cite{datta_2005}. This obviously holds for the delta function of a discrete eigenvalue, but it must also hold for states undergoing broadening; suppression of the DOS amplitude directly corresponds to an increased broadening, such that the integrated peak always sums to 1.

The LDOS of a state of given energy is related to the imaginary part of the diagonal elements of the Green's function associated with the Hamiltonian evaluated at the same energy \cite{datta_2005, economou2006green}:
\begin{equation}\label{ldospsi}
|\psi_n(E)|^2 \propto -\Im\{ \left(G_\text{SSH}^{\,\mathrm{eff}}(E)\right)_{nn}\}.
\end{equation} 

\begin{figure}
\centering\includegraphics[width=1\linewidth]{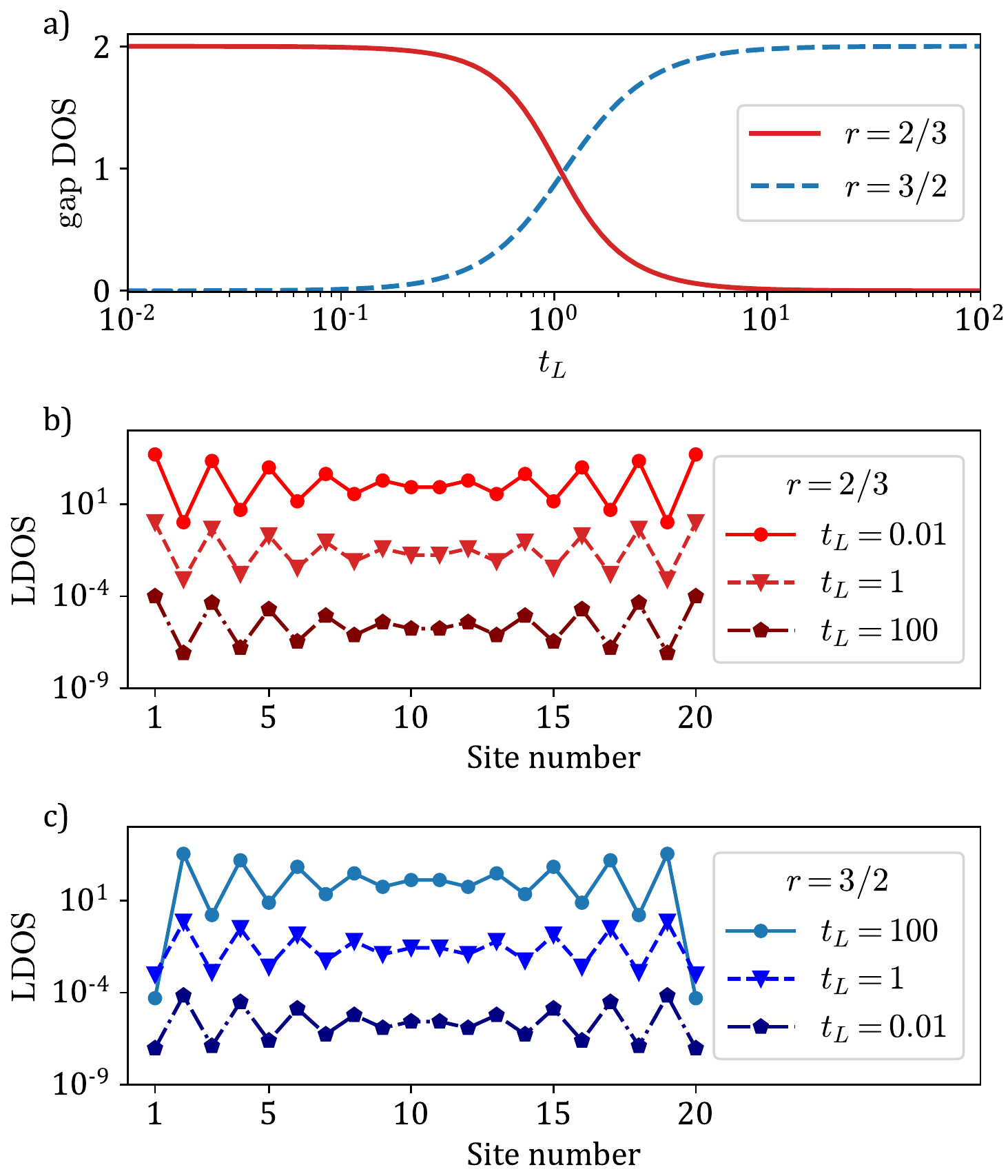}
    \caption{a) Integrated DOS over the SSH chain band gap in the effective system as a function of $t_L$, showing the response of low-energy states to the environmental coupling. The $r=2/3$ case (red) corresponds to a topological SSH chain with $(t_1,t_2)=(0.8,1.2)$, and shows the suppression of edge states. The $r=3/2$ case (blue) corresponds to a topologically trivial SSH chain with $(t_1,t_2)=(1.2,0.8)$ and shows the emergence of PIE states. b) Topological edge state LDOS suppression corresponding to the $r=2/3$ case. Note that for $t_L=1$ (on the order of SSH chain parameters), LDOS suppression spans nearly four orders of magnitude. c) PIE state LDOS growth in a trivial SSH chain, clearly showing that first and last sites lose support for large $t_L$ due to the formation of islets.
}
    \label{fig:gapdosfig}
\end{figure}

Making use of the DOS sum rule, integration over a chosen energy range allows one to count states of the effective system within that range. The integrated DOS over the SSH chain's band gap, $\pm\Delta=\pm|t_1-t_2|$, referred to as the gap DOS, is a constant if the number of in-gap states remains unchanged, so long as the broadening is smaller than the gap width. For broadening exceeding $\Delta$, as seen near $t_L=1$ in Fig.~\ref{figspp}, the gap effectively closes as a significant portion of the state leaks into the bands, leading to a decreasing gap DOS. A complete decay of the gap DOS in either the weak- or strong-coupling limit (where as has been argued above the SSH chain decouples from the rest of the system, losing two sites in the strong coupling limit), reflects the disappearance of states within the gap. The gap DOS is shown for both phases of the SSH chain in Fig.~\ref{fig:gapdosfig}a. 

Edge states in the topologically nontrivial SSH chain ($r<1$; red curve) remain intact for small values of $t_L$; a strong suppression of the gap DOS is observed around $t_L=1$. For $t_L>1$, the gap DOS tends to zero as $1/t_L^2$, indicating the loss of topological edge states. The topological edge state LDOS of the $r<1$ case is shown in Figure \ref{fig:gapdosfig}b. One observes a large LDOS for small $t_L$ which rapidly decreases for large $t_L$. This is consistent with the previous interpretations of the system in the limits of small and large coupling ($t_L\ll 1$ and $t_L \gg 1$, respectively). In the first limit, the system is essentially isolated such that the singularities of the effective Green's function $G^{\,\mathrm{eff}}_{\text{SSH}}$ lie near the real axis, demonstrating the large edge state LDOS. In the second limit, the remaining SSH chain is of length $N-2$, and edge states have vanishing  support on the restricted segment spanning  sites 2 through $N-1$.
This is reflected in the low-energy poles in $G^{\,\mathrm{eff}}_{\mathrm{SSH}}$ which acquire large imaginary parts, yielding an edge state LDOS near 0. The suppression of both edge state DOS and LDOS is well approximated as having a $1/t_L^2$ scaling, with minor deviations when $t_L\lesssim1$. 

For the topologically trivial SSH chain ($r>1$; \ref{fig:gapdosfig}a blue curve), no low-energy modes can be supported in an isolated chain for small $t_L$ where the gap DOS is effectively zero. As $t_L$ is increased, the gap DOS approximately scales as $t_L^2$ while $t_L<1$ and plateaus for strong couplings, revealing the emergence of PIE states. The emergence of PIE states is also reflected in the LDOS of the $r>1$ configuration, shown in Figure \ref{fig:gapdosfig}c. The LDOS and DOS of PIE states both approximately scale as $t_L^2$, with small deviations when $t_L\gtrsim1$. Unsurprisingly, for small $t_L$, PIE states have a small LDOS. In the limit of strong coupling, these low-energy states can be understood as edge states of a truncated SSH chain of length $N-2$; the Green's functions here has low-energy poles near the real axis, yielding a large LDOS. The formation of islets is further evidenced by the highly suppressed amplitudes on sites 1 and $N$ at strong coupling. 

The DOS and LDOS behaviors for $N$ odd are  markedly different due to the coexistence of edge and PIE states for either topological phase of the SSH chain. While for $N$ even edge states are suppressed ($r<1$) or PIE states emerge ($r>1$) with increasing $t_L$, the process is simultaneous in chains with $N$ odd for any $r$ and involves a delocalization-localization transition from the left (right) boundary to the right (left). Interestingly, this implies that an odd chain can appear to have a localized mode near both boundaries for intermediate values of $t_L$, in sharp contrast with an isolated chain.

\subsection{Disorder}
\begin{figure}[ht]
\centering
\includegraphics[width=1\linewidth]{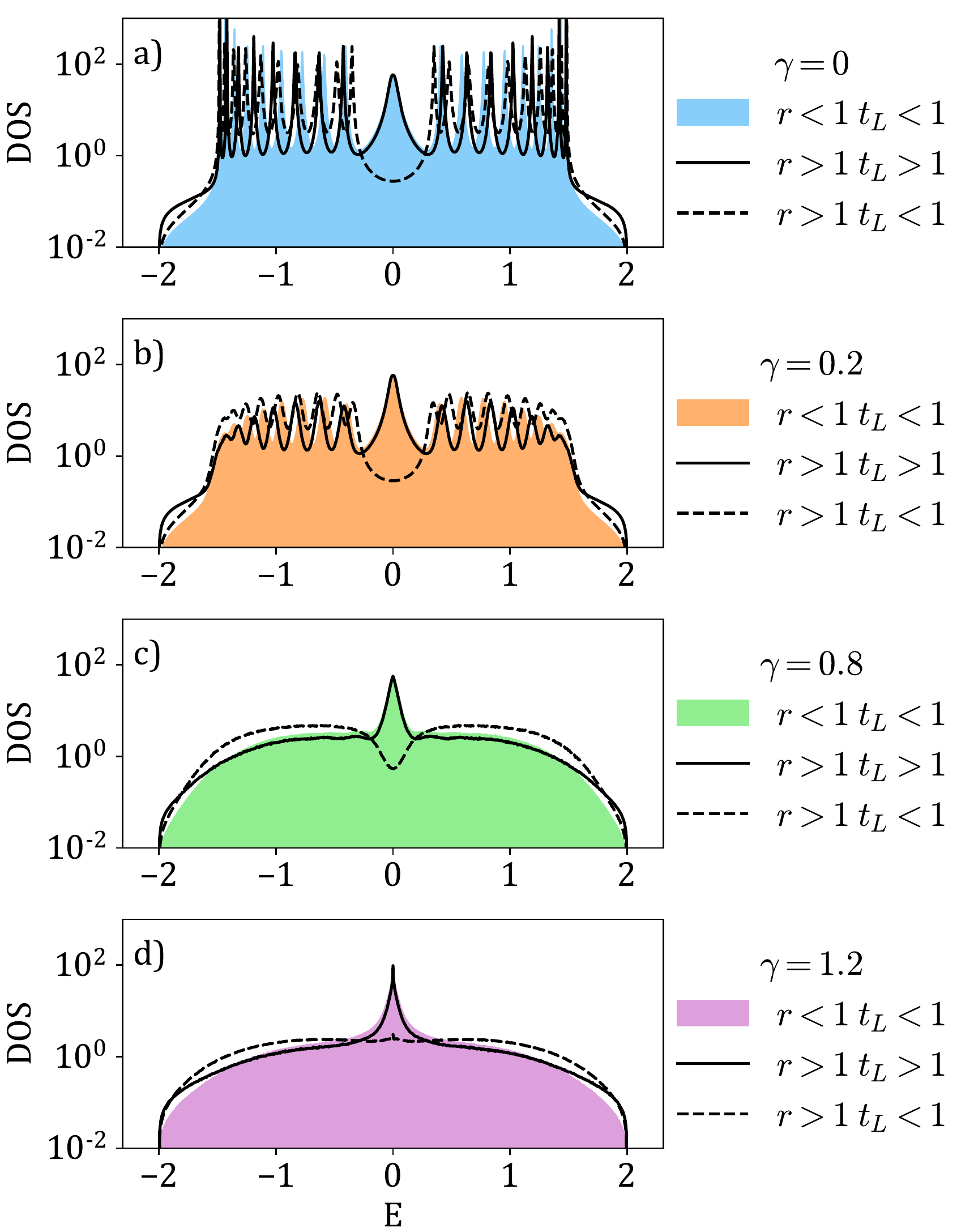}
\caption{DOS for several disorder strengths a) $\gamma=0$, b) $\gamma=0.2$, c) $\gamma=0.8$, and d) $\gamma=1.2$, with an SSH chain of length $N=20$. The conventional topologically nontrivial system ($r=2/3$ and $t_L=0.25$) is shown in shaded colors. Solid black lines denote the PIE states  for $r=3/2$ and $t_L=4$. Dashed black lines denote the trivial configuration for $r=3/2$ and $t_L=0.25$. A total of 10000 disorder configurations were considered and log-averaged.}
\label{disfig}
\end{figure}

An analysis of the influence of disorder on the edge and PIE states is carried out in the $N$-site effective representation $H^{\mathrm{eff}}_{\mathrm{SSH}}$. Unlike bulk states that are susceptible to disorder, topological modes are robust to certain symmetry-preserving forms of disorder \cite{2016,OSTAHIE,PhysRevB.101.144204}. The DOS is rendered as a function of the disorder strength parameter, $\gamma$, in Fig.~\ref{disfig} for  conventional edge states ($r<1$ and $t_L<1$; shaded color), PIE states ($r>1$ and $t_L>1$; solid black), and the trivial configuration for small coupling ($r>1$ and $t_L<1$; dashed black). Chiral disorder on individual $t_m$ elements (acting only on hopping parameters) is injected into the system in the form $t_m \rightarrow t_m (1+\delta)$, where $t_m$ is either $t_1$ or $t_2$, and where $\delta$ is sampled from a uniform random distribution $[\ -\gamma, \gamma]$.

Disordered densities of states for both edge and PIE states are similar, with the distinction that the total number of states at $\gamma=0$ is $N$ for $r<1$ and $t_L<1$ (edge state), and $N-2$ when $r>1$ and $t_L>1$ (PIE state), as expected due to islet formation. All states show broadening at $\gamma=0$ due to the non-zero (non-infinite) $t_L$ value chosen. The value of $t_L$ for both systems was chosen such that the low-energy mode broadening is comparable. Injecting disorder  introduces noise into the bulk bands, which, after configuration averaging leads to flat and featureless bands for large enough disorder. Note that the bulk bands close the gap at $\gamma=1$. 

Interestingly, both types of broadened low-energy states within the SSH chain subspace are robust to disorder and remain supported until $\gamma=1$. For $\gamma>1$,  low-energy modes remain but change from a broadened peak to a very high-DOS state arising at $E=0$, hinting that even in this gapless configuration some form of topological protection remains at the mid-gap. As expected, the trivial $r>1$ configuration shows no peak about $E=0$, although a small increase in the DOS is visible for stronger disorder ($\gamma>1$). Unsurprisingly, the conventional ($r<1$) SSH chain edge states are robust to disorder. This simple exercise has nonetheless confirmed the topologically robust nature of PIE states, which qualitatively behave  identically to conventional edge states.

\subsection{Transmission}
Further evidence of the topological transition as a function of the coupling $t_L$ arises when considering transmission. In the effective description, transmission can easily be written in terms of $\left(G^{\,\mathrm{eff}}_{SSH}(E)\right)_{N1}$, the corner element of the effective  Green's function in the SSH chain subspace:
\begin{equation}
    T=4\Im(\Sigma_\infty)^2|\left(G^{\,\mathrm{eff}}_{SSH}\right)_{N1}|^2.
\end{equation}

\begin{figure}[h]
\centering
\includegraphics[width=1\linewidth]{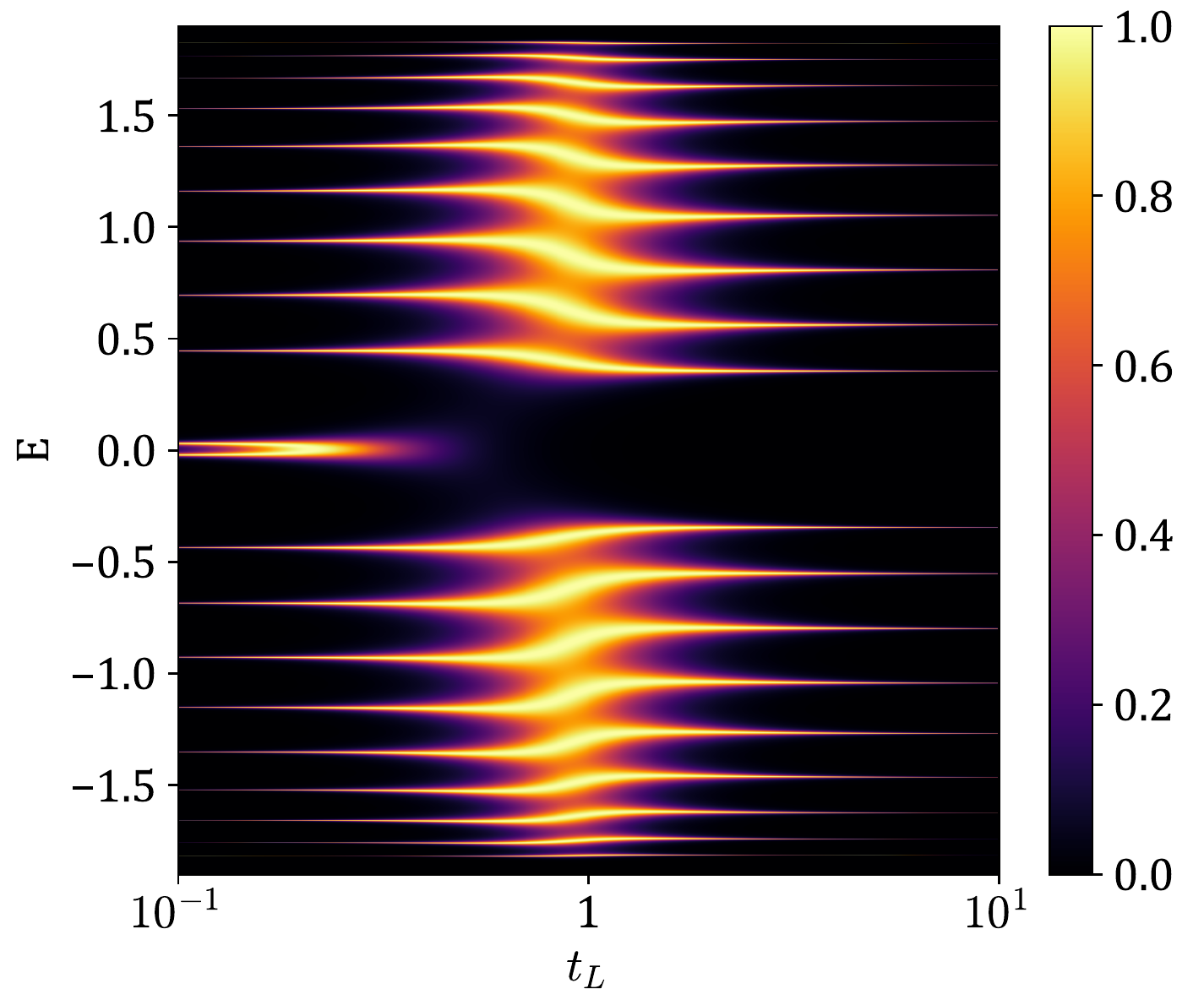}
\caption{Transmission through an SSH chain of length $N=20$ with $t_1=0.6$ and $t_2=0.8$ as a function of $t_L$. As the DOS becomes discrete for small and large couplings, the transmission channel width goes to zero in these limits.  Due to computational resolution, these channels do not extend to the margins of the plot.
}
\label{figT}
\end{figure}

Transmission can be computed over the entire energy spectrum using the energy-dependent self-energy $\Sigma_{\infty}(E)$. Naturally, states supported within the SSH subspace lead to unitary transmission, as they provide valid channels for transport. A typical transmission spectrum is shown in Fig.~\ref{figT}.
Bulk band transmission is sustained for all values of $t_L$. Shifts in the transmission channels correspond to the SSH bulk band levels evolving from those of an isolated SSH chain of length $N$ to those of a chain of length $N-2$.

\begin{figure}[ht]
\centering
\includegraphics[width=1\linewidth]{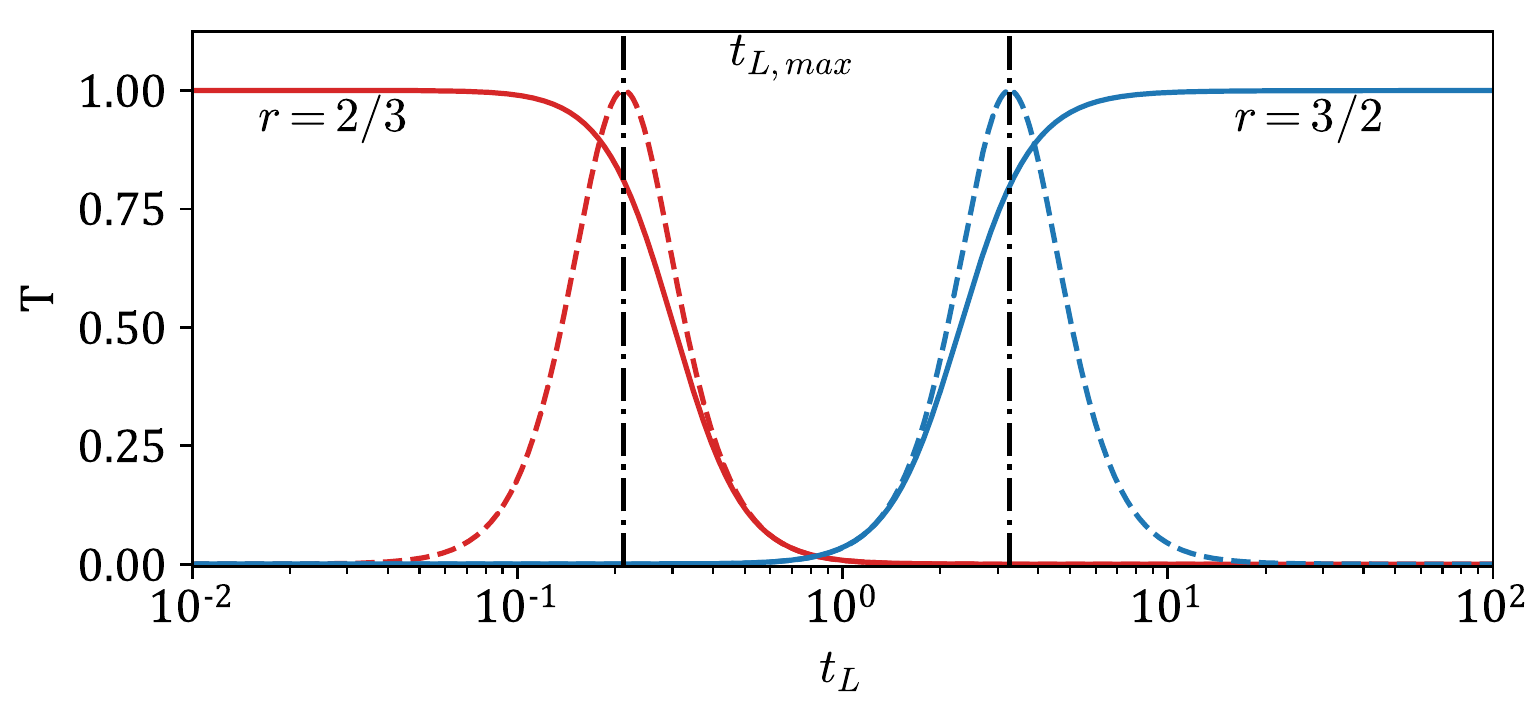}
\caption{Transmission of edge states (red) and PIE states (blue) as a function of $t_L$. For an SSH chain of $N=20$, $t_1=0.6$, $t_2=0.8$ ($r=2/3$), the transmission at the edge state energy is given by the solid red line while the dashed line shows the transmission at $E=0$. In the opposite phase, $t_1=0.8$ and $t_2=0.6$ ($r=3/2$), transmission at the PIE state energy is given by the solid blue line while the dashed line denotes $E=0$. The vertical lines label the unitary maxima for the $E=0$ cases in the two phases $r<1$ and $r>1$, respectively. }
\label{figTTES}
\end{figure}

The behavior of edge state transmission is strikingly different. The edge states show a unitary transmission plateau which merges at $E=0$ for $t_{L,\text{max}}=\sqrt{t_2}(t_1/t_2)^{N/4}$, found by evaluating $dT(E=0)/dt_L=0$. For $t_L>t_{L,\text{max}}$, transmission is  quickly suppressed. Studying the transmission at $E=0$, as was done in \cite{OSTAHIE}, reveals a non-monotonous dependence of transmission on $t_L$. If instead transmission is studied at the exact edge states energies, which vary as a function of $t_L$, one can see unitary transmission near $t_L=0$ as would be expected of edge states in a weakly-coupled system. A comparison of the two cases is shown in Fig.~\ref{figTTES}. This figure also shows the transmission associated with PIE states arising in the opposite topological phase.
This reveals that unitary transmission of edge states in these systems is highly dependent on both the exact topological edge state energy and the coupling strength to the leads. This strong suppression of the transmission in the gap suggests that edge states can no longer be considered good channels within the SSH chain subspace. This result qualitatively is in good agreement with the LDOS of a topological SSH chain in this composite system, which shows that topological edge states are strongly suppressed as a function of $t_L$. 

One can further show that for an SSH chain originally in a topologically trivial configuration ($r>1$), a unitary transmission plateau appears at edge state energies for $t_L>1$. Additionally, SSH chains of odd length show the same soft reversal of the topological phase as a function of $t_L$. There, rather than simply being suppressed, edge states delocalize from their original boundary to localize at the opposite boundary, as expected of a change of topology for odd-$N$ chains.

\section{Conclusion}
In this paper, we have analyzed a tripartite system composed of an SSH chain and two leads from different angles. In the first place, the system in its entirety was studied. It was shown that the wave functions associated with zero-energy states are given either by a linear combination of edge states or a linear combination of so-called PIE states. High-energy states were then studied, for which a transcendental equation was found for $E$; this allowed the computation of an energy spectrum for $E>2$. Using this spectrum, states with no analog in the closed SSH system were found: Tamm states. These two findings, in the limit of strong coupling, led to the notion of islet formation and SSH chain truncation, and first hinted to a topological phase transition.

An effective system, derived from the full tripartite system, was then studied. This effective approach encodes the full system in an effective Hamiltonian of a system of $N$ or $N-2$ sites, which are particularly useful for $t_L\ll1$ and $t_L\gg1$, respectively.

Using the $N$-site Hamiltonian, it was found that solutions outside of the lead band exactly match solutions of the tripartite system. Differences arise for $|E|<2$, where the effective system has a discrete spectrum of complex solutions while the full system has a continuous spectrum of real solutions. These differences simply reflect the subspace restriction within the effective system; a projection of the discrete spectrum onto the real axis restores the expected continuous spectrum. It was then demonstrated that the wave functions of the effective system are equivalent to those of the tripartite system within the SSH chain.

An $(N-2)$-site effective Hamiltonian was employed to examine low-energy states in the limit of strong coupling $t_L$. In this limit, the segment of the SSH chain going from site $2$ to site $N-1$ is essentially isolated from the environment. A topologically trivial SSH chain undergoing this truncation then supports PIE states, confirming that a topological phase transition occurs between $t_L=0$ and $t_L\rightarrow \infty$.

The behavior of edge states and PIE states was then studied for intermediate values of $t_L$ using the Green's function associated with the $N$-site effective Hamiltonian. Both the DOS and LDOS were studied in this regime, showing that low-energy modes are conventional edge states for $r<1$ while they are PIE states for $r>1$. The suppression (emergence) of edge states (PIE states) was  characterized as a function of the coupling $t_L$. A subsequent analysis of disorder and transmission further affirms the topological nature of PIE states.

Combining these results, the following picture emerges. The tripartite system has two types of low-energy states, edge states and PIE states, with support depending on the topological phase of the system and the coupling strength to the environment $t_L$. For $t_L$ near zero, the system has either zero or two low-energy topological edge states. In the former case, an increase of $t_L$ leads to PIE states emerging and tending towards being eigenstates of the $N-2$ SSH chain subspace. For the latter, the conventional edge states are suppressed and show decreased support within the subspace as $t_L$ is increased (their DOS tends to zero as $t_L\rightarrow \infty$). Furthermore, in the limit of strong coupling, Tamm states associated with the formation of independent islets emerge and the SSH chain undergoes truncation from $N$ sites to $N-2$ sites.

Due to the boundary-localized nature of the coupling $t_L$ studied in this work, no topological invariant as a function of $t_L$ could be found via standard bulk-invariant methods. To obtain invariants that include environmental influence, one needs to turn to open quantum (non-Hermitian) systems. To date, many non-Hermitian variants of the SSH Hamiltonian have been studied, including the addition of chiral gain and loss along hopping parameters \cite{PhysRevB.97.045106}, and alternating on-site gain and loss \cite{articleDangel,PhysRevB.97.121401}. Obtaining invariants is possible in such systems as the non-Hermitian terms are bulk properties of the models. Systems similar to the tripartite system studied here featuring boundary couplings have been studied \cite{articleDangel,OSTAHIE,articleMarques} but have so far evaded rigorous characterization via invariants due to the open and non-Bloch nature of these models. New ideas being developed may lead to an understanding of topological classifications for non-Bloch non-Hermitian systems in the near future \cite{Kunst, Leykam, Imura, Yokomizo}. 

This work highlights the importance of accounting for the influence of the environment in the development of topological devices. Careful considerations in design are necessary to ensure a device will exhibit the desired topological phase; something as simple as dissipative boundary couplings can induce a phase transition and cause edge states to vanish, or conversely cause PIE states to emerge. Other one-dimensional models hosting symmetry-protected topological phases in particular are expected to be susceptible to unwanted transitions induced from a local breaking of symmetry cause by boundary couplings.

\begin{acknowledgments}
This work was supported in part by the Natural Science and Engineering Research Council of Canada (\#RGPIN-2020-05094) and by the Fonds de Recherche Nature et Technologies du Québec via the INTRIQ strategic cluster grant (\#2018-RS-201919).
\end{acknowledgments}

\appendix*

\section{Energy spectrum at strong coupling}
\label{sec-appA}

The energies displayed for the full system in Fig.~\ref{spec_1} and for the effective system in Fig.~\ref{figspp} are easily understood for small $t_L$ simply by taking the limit $ t_L \to 0 $. In this limit the system splits up into three uncoupled, easily solved subsystems: the two semi-infinite leads and the SSH chain. Strong coupling is not quite so obvious, but perturbation theory can be used to shed light on the energy spectrum. We define a rescaled version of the Hamiltonian \eqref{eq1nic} by
\begin{equation} \label{eq-hprime}    
    \hat H \equiv \dfrac{1}{t_L} H = H_0 + \frac{1}{t_L} H_1,
\end{equation}
where 
\begin{equation} \label{eq-h0}
    H_0 = c_1^\dagger l_1 + l_1^\dagger c_1 +c_N^\dagger r_1 + r_1^\dagger c_N
\end{equation}
and
\begin{equation} \label{eq-h1}
    H_1 = H_{\text {L}} + H_{\text{SSH}} + H_{\text {R}}
\end{equation}
are independent of $t_L$. We now perform perturbation theory on $\hat H$, treating $1/t_L$ as the small parameter.

In matrix form, $H_0$ and $H_1$ are block diagonal:
\begin{align}
    \label{eq-blocks1}
        H_0 & = \text{diag} \left( 
        \mathbb{O}_\infty, 
        \sigma_1,
        \mathbb{O}_{N-2}, 
        \sigma_1,
        \mathbb{O}_\infty 
    \right),
    \\
     \label{eq-blocks2}
        H_1 & = \text{diag} \left( 
        H_L,
        H_{\text{SSH}},
        H_R 
    \right),
\end{align}
%
%
%
%
%
%
where $ \mathbb{O}_{j} $ is a zero matrix of dimension $ j $, $ \sigma_1 $ is the first Pauli matrix and the elements of \eqref{eq-blocks2} are the matrix equivalents of the corresponding elements of \eqref{eq-h1}.
Note that the blocks in \eqref{eq-blocks1} and \eqref{eq-blocks2} don't align: the blocks in \eqref{eq-blocks1} correspond to the elements in Fig.~\ref{islets} while those in \eqref{eq-blocks2} correspond to the elements in Fig.~\ref{f21}.

The unperturbed problem, with Hamiltonian $ H_0 $, has eigenvalues $ 0, \pm 1 $, each of which is degenerate. To perform degenerate perturbation theory, we first form a matrix whose columns are the eigenvectors of each degenerate subspace. For unperturbed energy $E_0 = \pm 1$, these matrices (corresponding to the Tamm states) are
\begin{equation} \label{eq-Psi-pm}
    \Psi_{\pm} = \dfrac{1}{ \sqrt{2} }
    \left( 
        \begin{array}{cc}
            \mathbf{0_\infty} & \mathbf{0_\infty} 
            \\
            \hline
            1 & 0 
            \\
            \pm 1 & 0
            \\
            \hline
            \mathbf{0}_{N - 2} & \mathbf{0}_{N - 2}
            \\
            \hline
            0 & \pm 1 
            \\
            0 & 1 
            \\
            \hline
            \mathbf{0_\infty} & \mathbf{0_\infty} 
        \end{array}
    \right),
\end{equation}
where $\mathbf{0}_{j} $ is a zero vector of dimension $j$ and the horizontal lines delineate the blocks of the unperturbed Hamiltonian (see \eqref{eq-blocks1}). For unperturbed energy $ E_0 = 0 $, the matrix (corresponding to all other states) is
\begin{equation} \label{eq-Psi-0}
    \Psi_0 
    = \left( 
        \begin{array}{ccc}
            \mathds{1}_{\infty} & \mathbb{O} & \mathbb{O} 
            \\
            \hline
            \mathbb{O} & \mathbb{O} & \mathbb{O} 
            \\
            \mathbb{O} & \mathbb{O} & \mathbb{O} 
            \\
            \hline
            \mathbb{O} & \mathds{1}_{N-2} & \mathbb{O} 
            \\ 
            \hline
            \mathbb{O} & \mathbb{O} & \mathbb{O} 
            \\
            \mathbb{O} & \mathbb{O} & \mathbb{O} 
            \\
            \hline
            \mathbb{O} & \mathbb{O} & \mathds{1}_{\infty}
        \end{array}
    \right).
\end{equation}
where $\mathds{1}_{j}$ is a $j$-dimensional unit matrix and $\mathbb{O}$ is a zero matrix whose dimension is determined by the other entries. The first-order corrections to  level $j$ are the eigenvalues of
the matrix
\begin{equation}
    \frac{1}{t_L} \Psi_j^\dagger H_1 \Psi_j.
\end{equation}
For $ E_0 = \pm 1 $ the corrections turn out to be zero while for $ E_0 = 0 $ the corrections are the eigenvalues of the matrix

\begin{equation}
    \frac{1}{t_L} \Psi_0^\dagger H_1 \Psi_0 = \frac{1}{t_L} \text{diag}\left( H_L,H_{\text{SSH}'},H_R \right),
\end{equation}
where $H_{\text{SSH}'}$ is the Hamiltonian of an SSH chain of length $N-2$ starting and ending with $t_2$. (We saw this Hamiltonian earlier; see Section \ref{sec-dog}.)

Combining the above results, the eigenvalues of $\hat H$ to first order in $1/t_L$ are $\pm1$ and $t_L^{-1}$ times the energies of the shortened SSH chain and two semi-infinite uncoupled leads. Multiplying by $t_L$, we obtain the eigenvalues of the Hamiltonian $H$ for large $t_L$. The spectrum consists of a pair of states at each of energies $E=\pm t_L$ (the Tamm states) and the energies of the isolated, shortened SSH chain and the leads, all with corrections of order $t_L^{-1}$. These are clearly seen at the right edge of Figs. \ref{spec_1} and \ref{figspp}.

\newpage
\bibliographystyle{apsrev4-1}
\bibliography{refsprobepaper}

\end{document}